# Extragalactic Objects in Basle Fields

**Selçuk BİLİR, Yüksel KARATAŞ, and Serap AK**

*İstanbul University, Science Faculty, Department of Astronomy and Space Sciences,*

*34452 Istanbul – Turkey*

sbilir@istanbul.edu.tr

**Abstract**

We discriminated the extragalactic objects (galaxies and quasars) from stars for Basle star fields, Plaut I, SA 54, and SA 82 by comparison of Basle fields and APS POSS I finder charts. Their numbers are 187, 70, and 93 for Plaut I, SA 54, and SA 82, respectively. Most of these objects are apparently faint in V, and they cause an overestimated local luminosity function for absolutely faint magnitudes, M(V) > 5. This effect is more conspicious for the field Plaut I.

**Key Words:** UBV photometry – Luminosity function – Extragalactic objects

1. Introduction

Modern star count works used to probe the structure of the Galaxy have been carried out for about two decades [1-9]. The effect of the contamination of extragalactic objects has been claimed by most of these authors. This effect becomes more important when apparently faint magnitudes are considered, where the extragalactic objects outnumber the stars [9].

Automated Plate Scanner catalog of the original Palomar Observatory Sky Survey (hereafter APS POSS I) provides us star/galaxy separation. Odewahn et al. [10, 11] address that an artificial neural network star/galaxy classifier has a success rate of 90% - 95% for galaxies with $18 \leq O \leq 20$ mag and a 98% success for stars in the same magnitude range.

The separation of the extragalactic objects has been omitted in the Basle Halo Program (BHP, [12]), probably due to relatively bright limiting magnitude, except in the work of Fenkart & Karaali [13] where seven images have been classified as "white objects" according to their position in the (U-G, G-R) two colour diagram. Though, we thought to use the APS finder charts of POSS I to identify the extragalactic objects in three Basle fields, Pault I, SA 54, and SA 82, to check the effect of these objects on the local luminosity function.

The outline of the paper is as follows: Data and the method are presented in Section 2. Magnitude and colour distribution for the extragalactic objects are discussed in Section 3. We mention the effect of extragalactic objects on the local luminosity function in Section 4. The conclusion is summarized in Section 5.

**2. Data and the Method**

The data and the finder charts for the fields Pault I, SA 54, and SA 82 are taken from the Basle Photometric Catalogues No: XII, VIII, and XIII, respectively [14-16]. The data for SA 54 and SA 82 are in UBV, whereas Plaut I has been investigated in RGU. We used the transformation equations of Buser [17] to transform the RGU data to UBV. Table 1 gives the coordinates, size, and the E(B-V) reddening for the fields.

APS catalogue of POSS I is from the scans of glass duplicated (blue and red) of the original Palomar Observatory [11]. The database includes position, magnitude (O and E mag), colour (O-E), and star/galaxy classification. The details about star/galaxy classifying technique are discussed in the works of Pennington et al. [11] and Odewahn et al. [10, 18]. The catalog of objects is available as on-line database over the Internet at the URL *http://aps.umn.edu/aps_catalog.html*. In addition to database searches on various catalog entry fields, one may produce finder charts on line.

We downloaded APS POSS I finder charts and the data for our fields in Table 1, by utilizing WEB page facilities of the Minnesota University. We compared the Basle field charts with the APS POSS I finder charts and revealed the extragalactic objects. Thus, we added the coordinates of the objects (stars or galaxies) and the star/galaxy identification to the catalogues of three Basle fields. The number of extragalactic objects are 187, 70, and 93 for the fields Plaut I, SA 54, and SA 82, respectively.

3. **Magnitude and Colour Distribution for Extragalactic Objects**

Fig.1 shows the position of the extragalactic objects in the (U-B) - (B-V) two colour diagrams. The curves are the iso-metallicity lines for the range of $-3 \leq [M/H] \leq +1$ dex, taken from Basle Stellar Library [19]. Fig.1 shows that the extragalactic objects do not only occupy the stellar regions but also they lie out of grid.

The V-magnitude distribution of the extragalactic objects are given in Table 2 and Fig.2. It is evident that these objects dominate the faint magnitudes. Their percentages relative to the total objects (stars and extragalactic objects) are 10%, 5%, and 10% for the fields Plaut I, SA 54, and SA 82, respectively. The normalized galaxy counts of Yasuda et al. [20] are also plotted in Fig.2. Their result is based on the Sloan Digital Sky Survey (SDSS) data obtained for the fields in the north and south Galactic caps with size 230 and 210 square-degree. There is a good agreement for the extragalactic object counts per magnitude per square-degree, between two works, for the intermediate latitude fields SA 54 and SA 82. The agreement is less for the low latitute field Plaut I ($b = +27^o.4$).

The colour distribution for the extragalactic objects are given in Table3, Fig.3 and Fig.4. These objects dominate the colour interval 0.5 < B-V < 1.0 for Plaut I, whereas they occupy a larger B-V interval for the fields SA 54 and SA 82, i.e. 0.5 < B-V < 1.5. One can notice a similar difference in the U-B distribution, i.e. the extragalactic objects occupy the interval U-B > 0.0 for the field Plaut I, whereas they are bluer, U-B > -0.5, for SA 54 and SA 82.

**4. Effect of the Extragalactic Objects on the Local Luminosity Function**

We drew the local luminosity function for all objects and for only stars, for three fields cited above to see the effect of the extragalactic objects on the luminosity function. We followed the procedure in our recent works [21-25]. The absolute magnitudes for dwarfs and sub-giants with [M/H] ≥ -1.75 dex are determined by the method of Laird et al. [26], whereas for extreme metal-poor stars, [M/H] < -1.75 dex, we used the Basle Stellar Library [19].

The result is given in Fig.5. The effect of the extragalactic objects is conspicuous for the field Plaut I (Fig.5a), where the number of these objects are larger than the numbers in the fields SA 54 and SA 82. However, the contamination of the extragalactic objects cannot be omitted for the last two fields, especially at the absolutely faint segment of the luminosity function.

## 5. Conclusion

In this paper, we have shown the contamination of the extragalactic objects on the local luminosity function for three Basle fields, for the first time. The effect of binarism and the evolved stars on the local luminosity function has been discussed in many of Basle works [21-25]. However, the contamination of the extragalactic objects has been omitted, probably due to relatively bright limiting magnitude. Here, we revealed that the extragalactic objects do effect even such works. Plaut I is the most conspicious among three fields, i.e. Plaut I, SA 54, and SA 82.

Meanwhile, we enriched the three Basle catalogues by identifying the extragalactic objects and adding the coordinates of the stars by comparison the Basle charts with the Minnesota ones.

**Acknowledgements:** We acknowledge financial support by Research Fund of the Istanbul University through Project No: 896/061296. Also, we would like to thank to the University of Minnesota for providing us with its APS on-line catalogues for the Basle fields in this study.

**Table 1.** The program fields.

| Field | α (1950) | δ (1950) | $l$ | $b$ | Size (□°) | E(B-V) |
|---|---|---|---|---|---|---|
| Plaut I | $16^h\ 06^m$ | $-13°\ 00'$ | $359°.3$ | $+27°.4$ | 0.54 | 0.07 |
| SA 54 | $10^h\ 26^m$ | $+29°\ 40'$ | $200°.1$ | $+58°.8$ | 2.56 | 0.00 |
| SA 82 | $14^h\ 16^m$ | $+15°\ 06'$ | $6°.3$ | $+66°.3$ | 1.20 | 0.00 |

**Table 2.** Magnitude distribution of the extragalactic objects and their percentage relative to the total number of stars and extragalactic objects.

| Field→ | Plaut I | | | SA 54 | | | SA 82 | | |
|---|---|---|---|---|---|---|---|---|---|
| V (mag) | Galaxy | Star | Percentage (%) | Galaxy | Star | Percentage (%) | Galaxy | Star | Percentage (%) |
| < 13 | 12 | 46 | 20.69 | 3 | 115 | 2.54 | 0 | 50 | 0.00 |
| 13-14 | 6 | 63 | 8.70 | 0 | 119 | 0.00 | 0 | 40 | 0.00 |
| 14-15 | 13 | 152 | 7.88 | 2 | 144 | 1.37 | 1 | 84 | 1.18 |
| 15-16 | 23 | 305 | 7.01 | 10 | 225 | 4.26 | 12 | 141 | 7.84 |
| 16-17 | 49 | 588 | 7.69 | 29 | 329 | 8.10 | 16 | 189 | 7.80 |
| 17-18 | 79 | 524 | 13.10 | 23 | 343 | 6.28 | 50 | 229 | 17.92 |
| 18-19 | 5 | 4 | 55.56 | 3 | 29 | 9.38 | 14 | 69 | 16.87 |
| **Total** | 187 | 1682 | 10.01 | 70 | 1304 | 5.09 | 93 | 802 | 10.39 |

**Table 3.** Colour distribution for the extragalactic objects for the fields Plaut I, SA 54, and SA 82.

| Field→ | Plaut I | SA 54 | SA 82 | Field→ | Plaut I | SA 54 | SA 82 |
|---|---|---|---|---|---|---|---|
| (B-V) | N | N | N | (U-B) | N | N | N |
| < 0.0 | -- | 2 | 3 | <-0.6 | -- | -- | 3 |
| 0.0-0.1 | -- | -- | 3 | -0.6-(-0.5) | -- | -- | 7 |
| 0.1-0.2 | 1 | 1 | 1 | -0.5-(-0.4) | -- | -- | 4 |
| 0.2-0.3 | 2 | 1 | 1 | -0.4-(-0.3) | 2 | 4 | 10 |
| 0.3-0.4 | 1 | -- | 5 | -0.3-(-0.2) | -- | 1 | 10 |
| 0.4-0.5 | 3 | 6 | 8 | -0.2-(-0.1) | 2 | 2 | 8 |
| 0.5-0.6 | 3 | 5 | 10 | -0.1-0.0 | 4 | 4 | 4 |
| 0.6-0.7 | 18 | 5 | 5 | 0.0-0.1 | 12 | 6 | 10 |
| 0.7-0.8 | 43 | 8 | 3 | 0.1-0.2 | 23 | 1 | 5 |
| 0.8-0.9 | 42 | 4 | 10 | 0.2-0.3 | 26 | 4 | 2 |
| 0.9-1.0 | 25 | 8 | 8 | 0.3-0.4 | 20 | 5 | 3 |
| 1.0-1.1 | 12 | 7 | 3 | 0.4-0.5 | 13 | 1 | 2 |
| 1.1-1.2 | 7 | 4 | 2 | 0.5-0.6 | 19 | 5 | 5 |
| 1.2-1.3 | 8 | 2 | 7 | 0.6-0.7 | 9 | 4 | 2 |
| 1.3-1.4 | 5 | 11 | 5 | 0.7-0.8 | 10 | 5 | 1 |
| 1.4-1.5 | 7 | 4 | 8 | 0.8-0.9 | 7 | 6 | 1 |
| 1.5-1.6 | 3 | 1 | 5 | 0.9-1.0 | 7 | 5 | -- |
| 1.6-1.7 | 1 | 1 | 5 | 1.0-1.1 | 11 | 2 | -- |
| 1.7-1.8 | -- | -- | -- | 1.1-1.2 | 8 | 2 | -- |
| 1.8-1.9 | -- | -- | -- | 1.2-1.3 | 4 | 2 | -- |
| 1.9-2.0 | -- | -- | 1 | 1.3-1.4 | 1 | 1 | -- |
| 2.0-2.1 | 1 | -- | -- | >1.4 | 9 | 10 | 16 |
| > 2.1 | 5 | -- | -- |  |  |  |  |
| **Total** | 187 | 70 | 93 | **Total** | 187 | 70 | 93 |

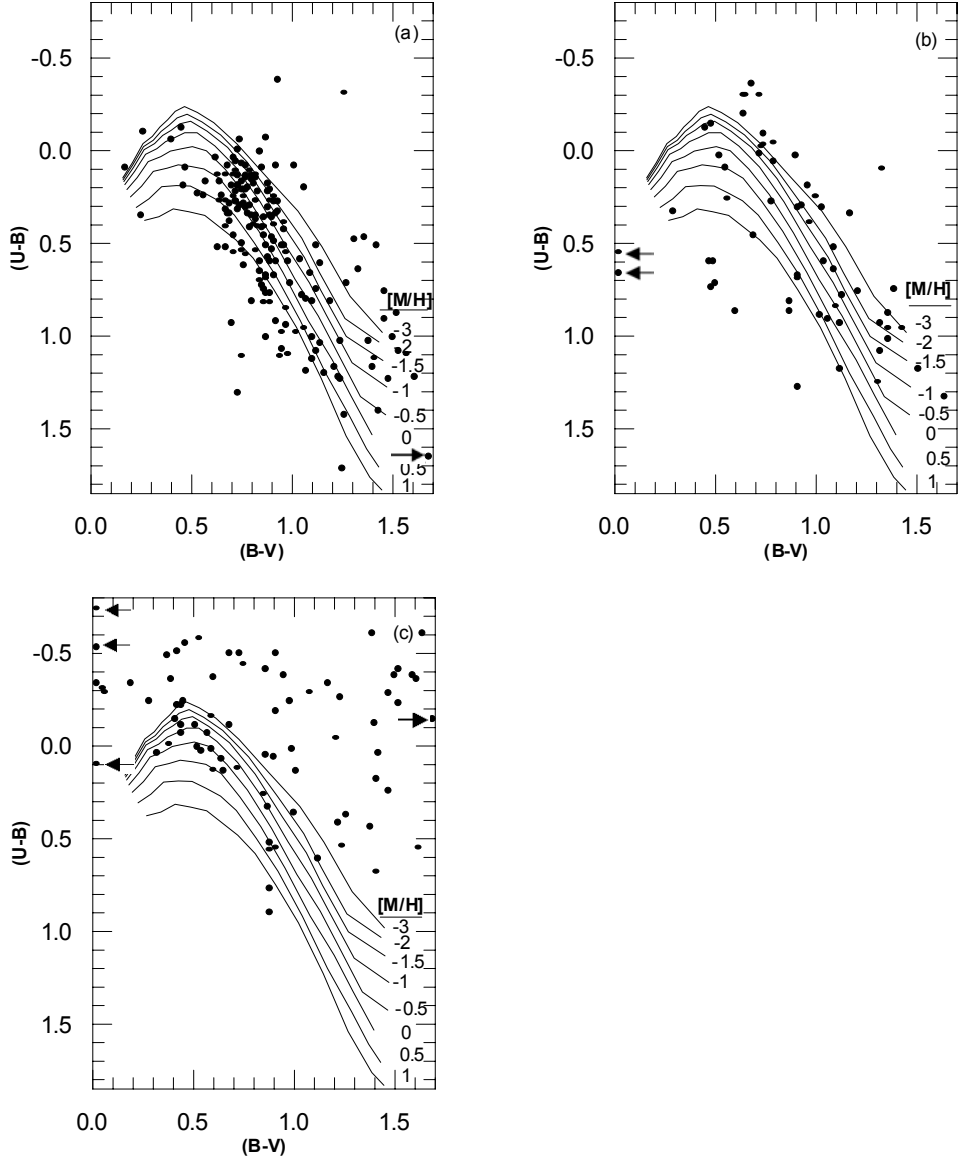

**Fig. 1.** Distributions of the extragalactic objects in the (U-B, B-V) two colour diagrams for three fields: (a) Plaut I, (b) SA 54, and (c) SA 82. Curves are the iso-metallicity lines for -3 ≤ [M/H] ≤ +1 dex.

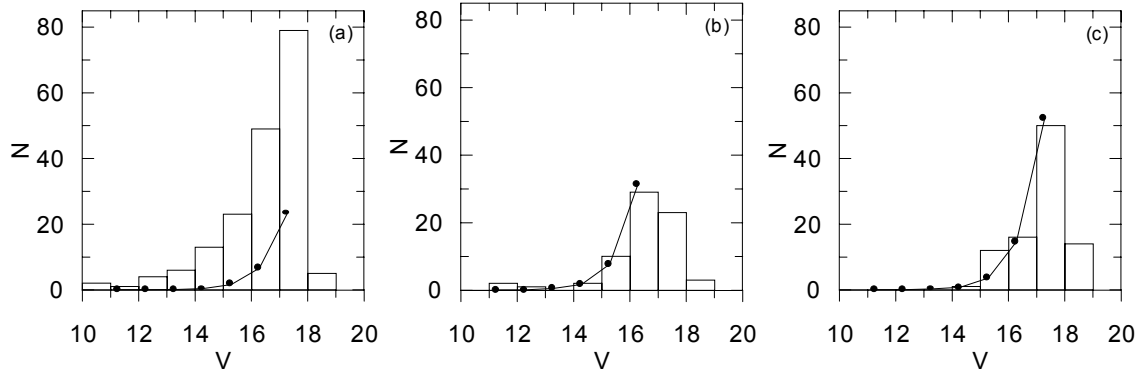

**Fig.2.** Apparent V magnitude distributions for the extragalactic objects in three fields: (a) Plaut I, (b) SA 54, and (c) SA 82. The normalized galaxy counts 1 mag deg$^{-2}$ for the north Galactic pole, for the SDSS data is shown as a solid line. Note the substantially larger impact of galaxies at faint magnitudes.

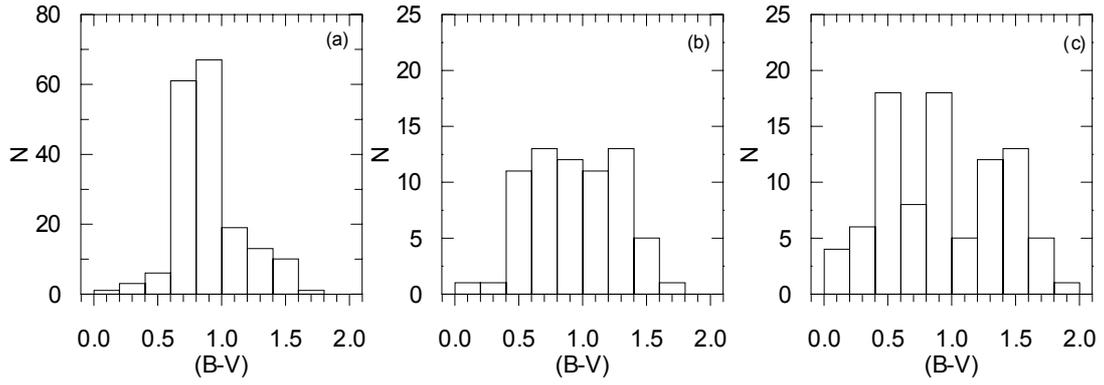

**Fig.3.** (B-V) colour distribution for the extragalactic objects for three fields: (a) Plaut I, (b) SA 54, and (c) SA 82.

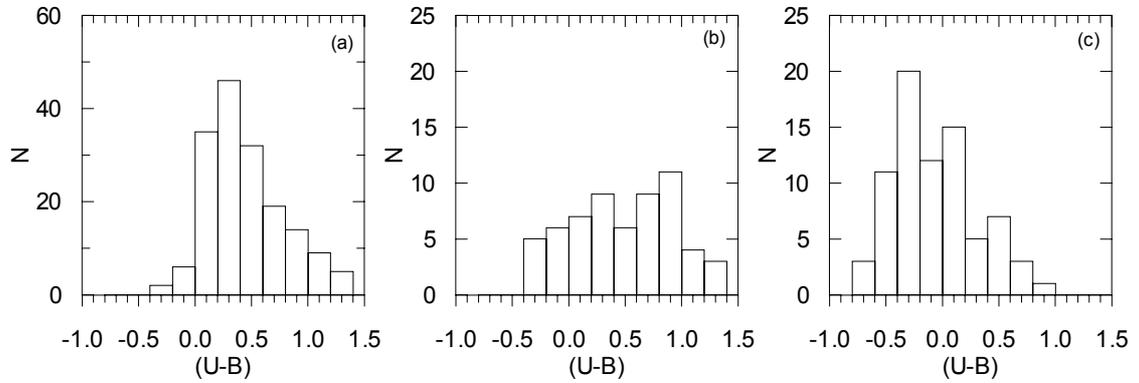

**Fig.4.** (U-B) colour distribution for the extragalactic objects for three fields: (a) Plaut I, (b) SA 54, and (c) SA 82.

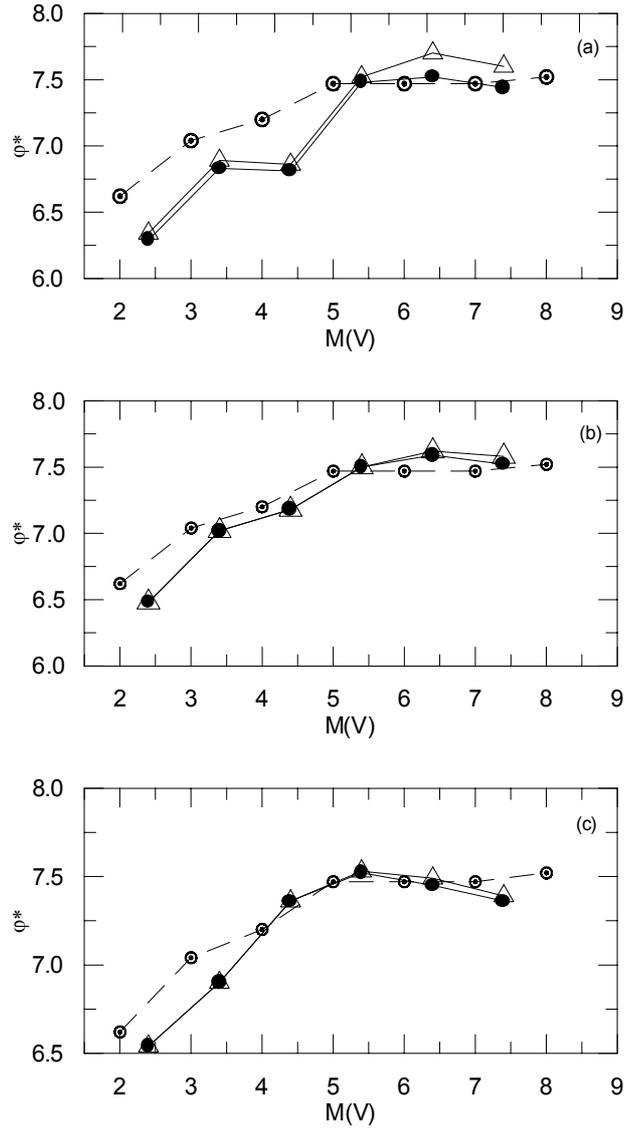

**Fig.5.** Effect of the extragalactic objects on the local luminosity function for three fields: (a) Plaut I, (b) SA 54, and (c) SA 82. Symbols: (⊙) Hipparcos [27], (Δ): stars and extragalactic objects, (●): only stars.